\documentclass[a4paper,11pt]{article}
\pdfoutput=1 

\usepackage{jinstpub} 
\usepackage{lineno}

\title{\boldmath Simulated Response of MuTe, a Hybrid Muon Telescope}

\author[a]{A. V\'{a}squez-Ram\'{i}rez$^1$,
}
\author[b]{M. Su\'{a}rez-Dur\'{a}n,}
\author[a]{A. Jaimes-Motta,}
\author[c,d]{R. Calder\'{o}n-Ardila,}
\author[a]{J. Pe\~{n}a-Rodr\'{i}guez,}
\author[e]{J. S\'{a}nchez-Villafrades,}
\author[a]{J.D. Sanabria-G\'{o}mez,}
\author[c,f,g]{H. Asorey,}
\author[a,h]{and L.A. N\'{u}\~{n}ez.}

\affiliation[a]{Escuela de F\'{i}sica, Universidad Industrial de Santander, Bucaramanga, Colombia}
\affiliation[b]{Departamento de F\'{\i}sica y Geolog\'{\i}a, Universidad de Pamplona, Pamplona, Colombia}
\affiliation[c]{Instituto de Tecnolog\'{i}as en Detecci\'{o}n y Astropart\'{i}culas, Centro At\'{o}mico Constituyentes, Comisi\'{o}n Nacional de Energ\'{i}a At\'{o}mica, Buenos Aires, Argentina}
\affiliation[d]{Instituto SABATO, Universidad Nacional de San Mart\'{i}n, Buenos Aires, Argentina}
\affiliation[e]{Escuela de Ingenier\'{i}a El\'{e}ctrica, Electr\'{o}nica y de Telecomunicaciones, Universidad Industrial de Santander, Bucaramanga, Colombia}
\affiliation[f]{Centro At\'{o}mico Bariloche, Comisi\'{o}n Nacional de Energ\'{i}a At\'{o}mica, San Carlos de Bariloche, Argentina}
\affiliation[g]{Escuela de Producci\'{o}n, Tecnolog\'{i}a y Medio Ambiente, Universidad Nacional de R\'{i}o Negro, \\San Carlos de Bariloche, Argentina}
\affiliation[h]{Departamento de F\'{i}sica, Universidad de Los Andes, M\'{e}rida, Venezuela}

\abstract{
In this paper we present a complete and detailed computational model of the response of the hybrid Muon Telescope (\textsl{MuTe}), designed to perform muography volcanic studies. This instrument combines two particle detection techniques: first, a muon hodoscope based on two panels of plastic scintillator bars; and a Water Cherenkov detector located behind the rear scintillator panel acting both as a coincidence and a discriminating detector. The simulation model includes: materials, geometries, dimensions, and the photo-sensitiveness of the detectors. The simulation results, in agreement with the measured data, were used to set up the muon detector trigger for the expected energy dependent signal.}


\arxivnumber{1912.10081} 

\begin{document}
\maketitle
\flushbottom

\section{Introduction}\label{sec:introduction}
Cosmic rays (\textsl{CR}) are continuously impinging on the Earth's atmosphere producing cascades of secondary particles called extensive air showers (\textsl{EAS}), having three main components:  the electromagnetic, the hadronic, and the muonic. The electromagnetic component produced by electrons, positrons, and photons, couple through several processes.  The hadronic component generated by barions and mesons originated through QCD effects and finally, the muonic component, due mainly to muons coming from the decay of charged pions, kaons, and other mesons through weak interaction processes. The energy of these muons comprises a broad spectrum, but only those with the highest energies ($>1$ TeV, produced during the very first interactions of the evolving particles cascade such as the decay of charmed mesons) are collectively known as prompt muons and have enough energy to cross from hundreds to thousand meters of rock\,\cite{MarteauEtal2012}.  

Muography is an emerging technique based on measuring the attenuation of the directional muon flux moving traversing geological or anthropic structures\,\cite{Kaiser2019}. Nowadays we are witnessing several new successful academic and commercial applications such as the detection of hidden materials in containers\,\cite{BlanpiedEtal2015}, archaeological building scanning\,\cite{MorishimaEtal2017, GomezEtal2016}, nuclear plant inspection\,\cite{FujiiEtal2013}, nuclear waste monitoring, underground cavities\,\cite{SaracinoEtal2017}, the overburden of railway tunnels\,\cite{ThompsonEtal2019} and vulcanology applications (see e.g., \cite{TanakaOlah2019} and references therein). The existence of more than a dozen active volcanoes in Colombia, which represents significant risks to the nearby population\,\cite{Cortes2016, Agudelo2016, Munoz2017}, motivated local research groups to explore possible applications of the muography technique\,\cite{AsoreyEtal2017B, SierraPortaEtal2018, PenaRodriguezEtal2018, GuerreroEtal2019, ParraAvila2019, PenarodriguezEtal2019}.  

Hodoscopes are the most common detectors designed and used for volcano muography and consist of two or more panels devised to identify muon trajectories. Projects like \textsl{MU-RAY}\,\cite{AnastasioEtal2013}, \textsl{ToMuVol}\,\cite{CarloganuEtal2013}, and \textsl{DIAPHANE}\,\cite{LesparreEtal2010} use hodoscopes based on different detection technologies: emulsion plates, resistive plate chambers, micromegas, multi-wire proportional chambers, and scintillators, just to mention the most common ones. Each of these techniques has advantages and disadvantages:  emulsion plate detectors\,\cite{MorishimaEtal2017, Nagamine2016} provide an excellent spatial resolution of the order of sub-microns, are passive, and easy to handle. On the other hand, they have short lifetimes, and it is not possible to discriminate the time-stamp of dynamic phenomena, because the recorded events accumulate in the plates. Gaseous detectors,  Resistive Plate Chambers\,\cite{SehgalEtal2016, Fehr2012}, Micromegas\,\cite{BouteilleEtal2016}, and Multi-Wire Proportional Chambers\,\cite{OlahEtal2018}, make possible the capture of the short traces due to the detected particles in the micrometre range. Finally, scintillation based detectors, using segmented\,\cite{FujiiEtal2013, LesparreEtal2012, TanakaEtal2009} or continuous scintillators\,\cite{NagamineEtal1995, AguiarEtal2015, TangEtal2016} are more robust and affordable than gaseous and emulsion detectors. Nevertheless, their spatial precision is not as good as the accuracy of the other detectors, since the segments generally are of the order of centimetres.

In this paper, we present a detailed computational \textsl{Geant4} \cite{AgostinelliEtal2003} model of the hybrid Muon Telescope (\textsl{MuTe}) compared with the first data taken in Bucaramanga-Colombia. \textsl{MuTe} combines two techniques: a hodoscope with two detection planes of plastic scintillator bars, and a Water Cherenkov Detector (\textsl{WCD}) which acts as an absorbing element and as a third active coincidence detector. The model includes materials, detailed geometries and dimensions, and the detector photo-sensitive device characteristics. In the next section, we briefly describe the rationale behind the \textsl{MuTe} design. Section \ref{sec:hodoscope-response} discusses the response from the scintillator bar hodoscope to the impinging cosmic ray background, with emphasis on the bar scintillator model and the possible attenuation effects. In section \ref{attenuation}, we compare the simulation results with data obtained from an experimental laboratory setup. The relationship between temperature and the breakdown voltage of the silicon photomultiplier (\textsl{SiPM}) used in the scintillator panels is shown in section \ref{sec:bar-data}. Section \ref{sec:wcd-response} presents the response of the \textsl{WCD} and also compares some of the results with recent lab measurements. Finally in section \ref{sec:conclusions} we compile some final remarks and conclusions.
\section{The \textsl{MuTe} instrumental design}
\begin{figure}
    \centering
    \includegraphics[scale=0.35]{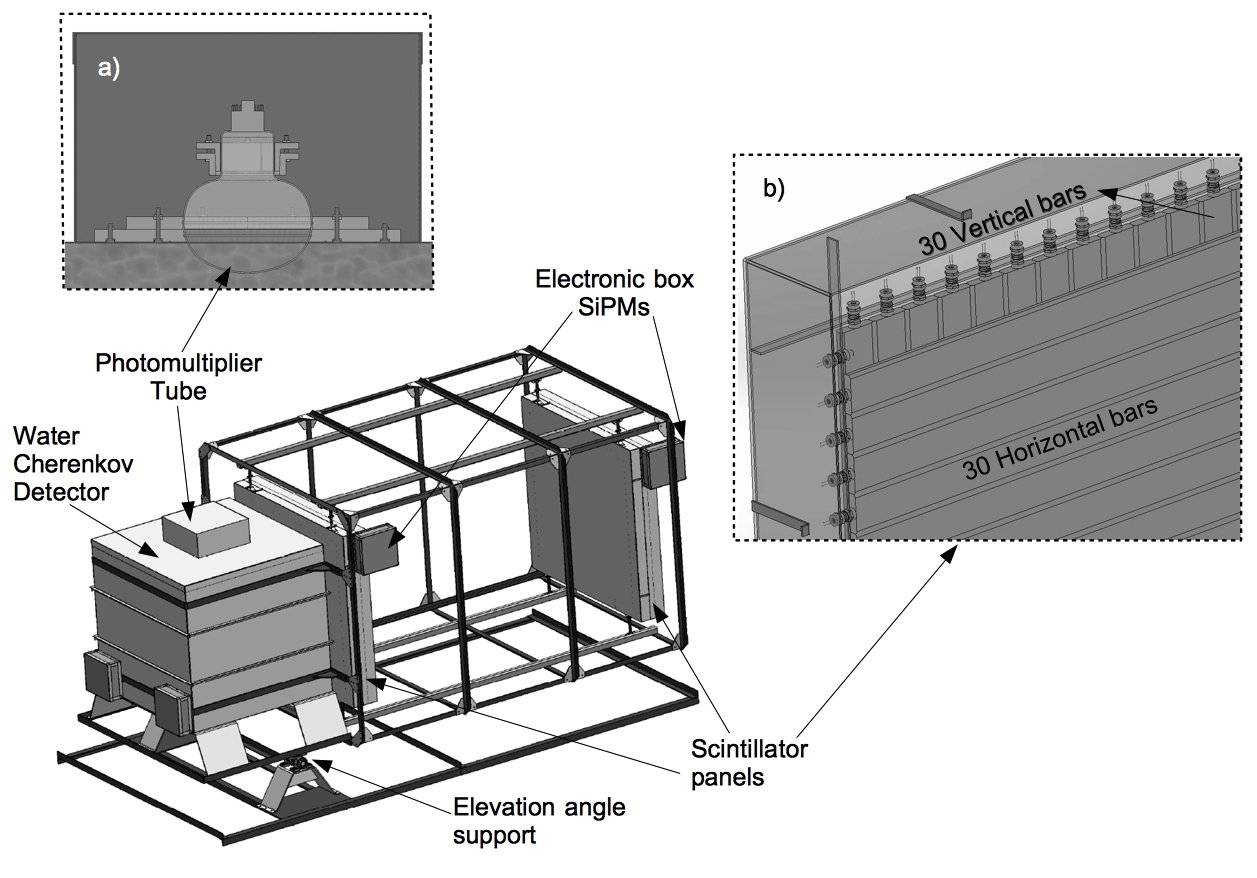}
    \caption{A sketch of the hybrid detector \textsl{MuTe}: a two-panel scintillator bar hodoscope in front and a Water Cherenkov Detector at the back. Each panel has X-Y  arrays of $30 \times 30$ plastic scintillating strips ($120$~cm $\times$ $4$~cm $\times$ $1$~cm), made with Styron$^{\textrm{TM}}$ 665-W polystyrene doped with a mixture of liquid organic scintillators: 1\% of 2,5-diphenyloxazole (PPO) and 0.03\% of 1,4-bis (5-phenyloxazol-2-yl) benzene (POPOP). Each array has $900$ pixels of $4$~cm $\times$ $4$~cm = $16$~cm$^2$, which sums up $14,400$~cm$^2$ of detection surface which can be separated up to a distance of $250$~cm. The \textsl{WCD} is a stainless steel container of $120$~cm side, coated inside with Tyvek, having a photomultiplier tube at the centre of the roof, with its sensitive surface in direct contact with the water as shown in plate \textit{a}. The \textsl{WCD} filters most of the upward background (caused by a flux of upward going particles) of muography and acts as a discriminator between the muonic and the electromagnetic part of the cascade. The \textsl{MuTe} mechanical structure has a variable elevation angle to adjust the telescope according to the object under study, and it is possible to adjust the scintillator panels separation to modify the pixel spatial resolution.} 
    \label{fig:mute-detector}
\end{figure}
The Muon Telescope shown in figure \ref{fig:mute-detector}, is a hybrid detector that combines two technologies: a two-panel scintillator bar hodoscope, and a water Cherenkov detector, devised to increase the signal/background ratio and to minimize the upward background\,\cite{NishiyamaMiyamotoNaganawa2014, KusagayaTanaka2015, NishiyamaEtal2016, GomezEtal2017}. This hybrid technique allows us to estimate not only the incoming flux directions but also the range of energy deposited by the impinging particles\,\cite{AsoreyEtal2017B, PenarodriguezEtal2019}.

The panels of the hodoscope consist of an array of $30$ vertical $\times\, 30$ horizontal scintillator bars (see plate \textit{b} in figure \ref{fig:mute-detector}), with dimensions $4$~cm $\times$ $1$~cm $\times$ $120$~cm long, providing $900$ pixels of $4$~cm $\times\,4$~cm, and a total detection surface of $14.400$~cm$^2$.

Each bar has a tunnel longitudinally drilled, where is inserted a Saint Gobain BCF92 multi-cladded wavelength shifter optical fiber (\textsl{WLS})\, \cite{SaintGobain2017}. Scintillation photons produced by the passage of charged particles through the bars are partially collected, absorbed, re-emitted in a different wavelength, and transported along the fiber. One of the ends of the fiber is polished at an angle of $45^{\circ}$ to increase internal reflection and to favour photon collection in the opposite end. Mechanically coupled to the fiber, is a silicon photomultiplier (Hamamatsu S$13360$-$1350$CS) consisting of $2668$ avalanche photodiodes\,\cite{Hamamatsu2018}. This device has a spectral detection range from $270$~nm to $900$~nm, with its maximum sensitivity around $450$~nm.

Casual coincidences detection and discrimination of the muon signals from upward background are enhanced by using an in-house developed Time-of-Flight (\textsl{ToF}) recording system, complemented by a cubic $120$~cm side water Cherenkov detector. The \textsl{WCD} consists of a cubic stainless steel container, internally coated with a reflective and diffusive material ($0.4$~mm Tyvek\,\cite{Filevich1999}) and filled with $1.7$~m$^3$ of purified water, having an $8$'' Hamamatsu R5912 photomultiplier tube (\textsl{PMT}) placed at the centre of the tank roof as shown in plate \textit{a}, figure \ref{fig:mute-detector}. The photosensitive \textsl{PMT} window is in direct contact with the water and can detect Cherenkov photons.


\section{Modelling the \textsl{MuTe} hodoscope }
\label{sec:hodoscope-response}
The response of the hodoscope bars refers to the signal produced in the \textsl{SiPM} photo-sensor when a charged particle crosses each scintillator bar. Indirect detection of secondary \textsl{EAS} photons is also possible mainly through Compton scattering and pair production. However, as the scintillation detector volume is relatively small, these mechanisms are highly improbable. 

\begin{figure}
    \centering
        \includegraphics[scale=0.25]{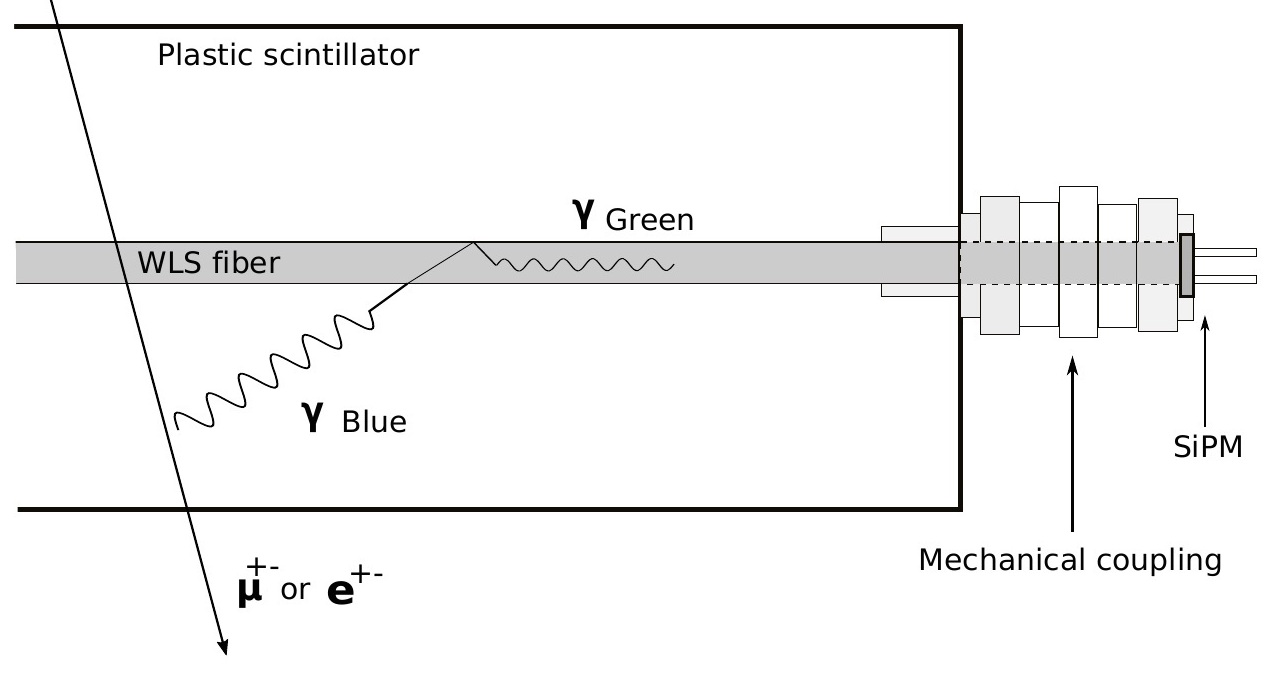}
   \caption{This sketch represents the scintillator bar system, with an embedded \textsl{WLS} fiber placed in the centre of the plastic scintillation bar and coupled to a \textsl{SiPM}. Charged particles crossing the bar produce scintillation photons in the blue-violet band. Those photons are absorbed and re-emitted in the \textsl{WLS} as green photons. The fiber guides photons to the \textsl{SiPM} where they can produce a signal depending on their wavelength, as the SiPM spectral detection range ($270$~nm to $900$~nm) has its the maximum sensitivity around $450$~nm.}
   \label{esquema_centelladora}
\end{figure}

As shown in figure \ref{esquema_centelladora}, charged particles impinging on the scintillators produce photons in the blue-violet-ultra violet bands, which can be absorbed by the plastic material or collected by the \textsl{WLS}. The fiber cladding captures the photons, where they can be absorbed, re-emitted (in the green optical band) and guided to the photo-sensor device located at one end of the fiber \cite{SaintGobain2017}. 

In the following sections we shall discuss the \textsl{Geant4} bar simulation model, the estimation of the average response of the hodoscope panels to the passage of charged particles --especially in the energy range of the Muon Reference Signal (\textsl{MRS}) which, for the present work will be of $3$~GeV-- and the experimental setup to measure the attenuation of the \textsl{Bar}-\textsl{fiber}-\textsl{SiPM} signal (see section \ref{sec:bar-data})

All the simulations performed are for muons with $3$~GeV, which are the most frequent at the $959$ m.a.s.l of Bucaramanga.  The Geant4 (version 10.3.2) modelling for the \textsl{MuTe} scintillator bar and \textsl{WCD} uses the \textit{QGSP\_BERT\_HP} physics list. The bar model, based on the \textsl{WLS} optical example, has an extra \textit{Optical Physics List} that includes the Wavelength Shifter, Cherenkov, Scintillation, Boundary effects and Absorption process, as well as the Rayleigh scattering and the Mie HG Scattering process, to simulate the propagation of photons inside a \textsl{WLS}  fiber. 

\subsection{The scintillator bar model} 
\label{sec:bar-simulation}
\begin{figure}
    \centering
    \includegraphics[scale=0.25]{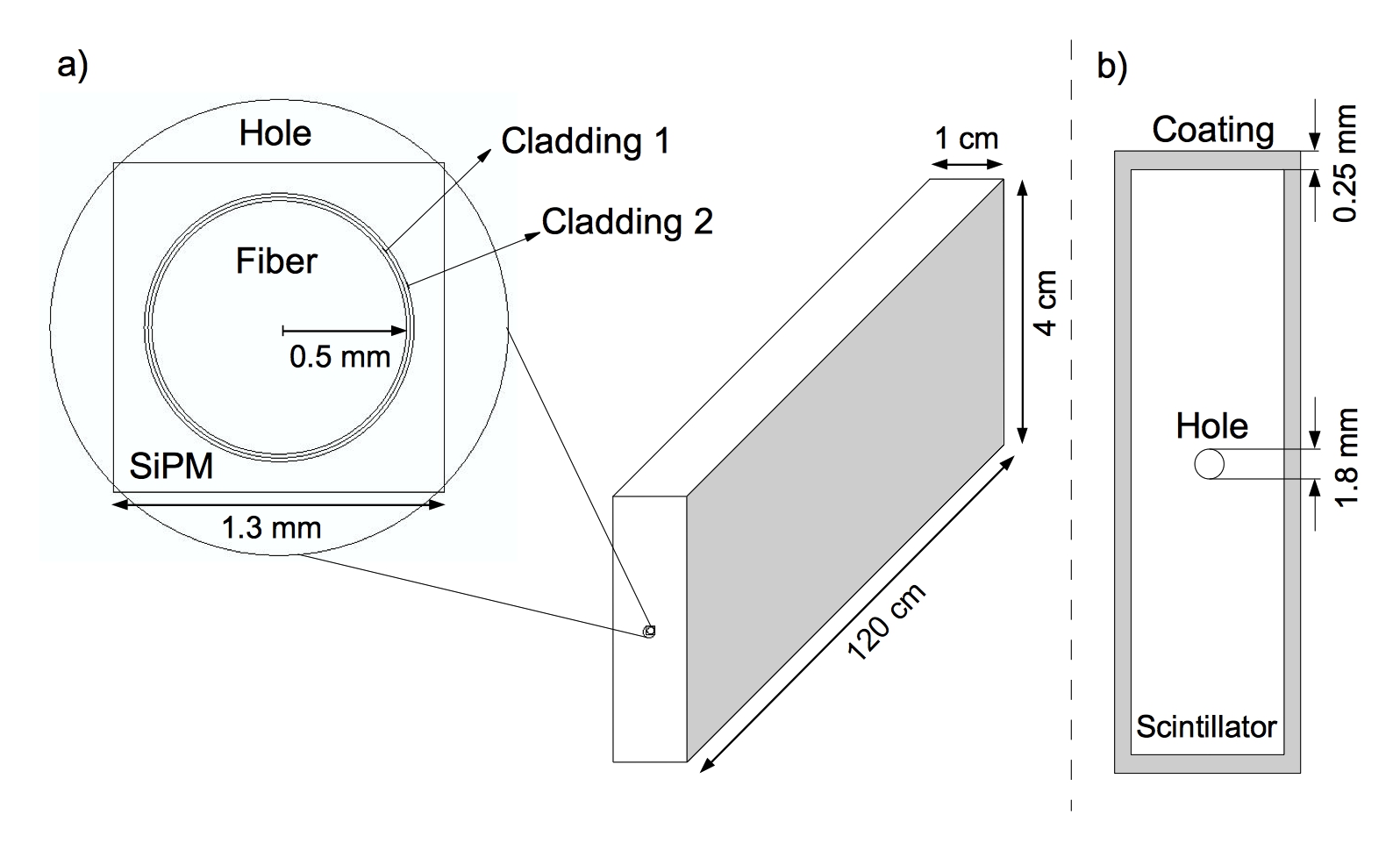}
    \caption{Geant4 model of the \textsl{MuTe} scintillator bar. a) The zoom of the hole shows the fiber, the double cladding and the SiPM dimensions that are attached to one end of the fiber. b) The scintillator has a coating material made of $15$\% TiO$_2$ and $85$\% polystyrene.}
    \label{fig:bar_dimentions}
\end{figure}
The \textsl{Geant4} model of the bar, shown in figure \ref{fig:bar_dimentions}, consist of a polystyrene scintillation parallelepiped with an index of refraction $n=1.50$, and a photon absorption length of $5.5$~cm. This bar has a coating material made of $15$\% TiO$_2$ and $85$\% polystyrene, with a reflectivity of $1$ and a thickness of $0.25$ mm (see figure \ref{fig:bar_dimentions} plate \textit{b}). The optical surface between the coating and the scintillator bar is a \textit{Glisur} model \textit{dielectric-dielectric} interface defined in \textsl{Geant4} with \textit{Ground finish} and polishment $
~=~1$. There is a tunnel ($1.8$~mm diameter), drilled through the central axis of the bar, in which a multi-cladded \textsl{WLS} fiber is inserted. As shown in plate \textit{a}, figure \ref{fig:bar_dimentions}, the fiber model is a solid cylinder of poly-methyl methacrylate, $119.45$~cm long and $0.5$~mm radius. The first cladding is a cylindrical shell of polyethylene $0.015$~mm thick and the second one is of fluorinated polyethylene with the same thickness. The optical surfaces between the fiber and the first cladding, and between the claddings are \textit{Glisur} models with \textit{Ground finish} \textit{dielectric-dielectric} interface, polishment and reflectance $=1$.

We model the \textsl{SiPM} as a square surface of side $1.3$~mm attached to one end of the fiber. The simulation employs the photon detection efficiency of the \textsl{SiPM} Hamamatsu S$13360$-$1350$CS installed in MuTe.

\begin{figure}
    \centering
    \includegraphics[scale=0.25]{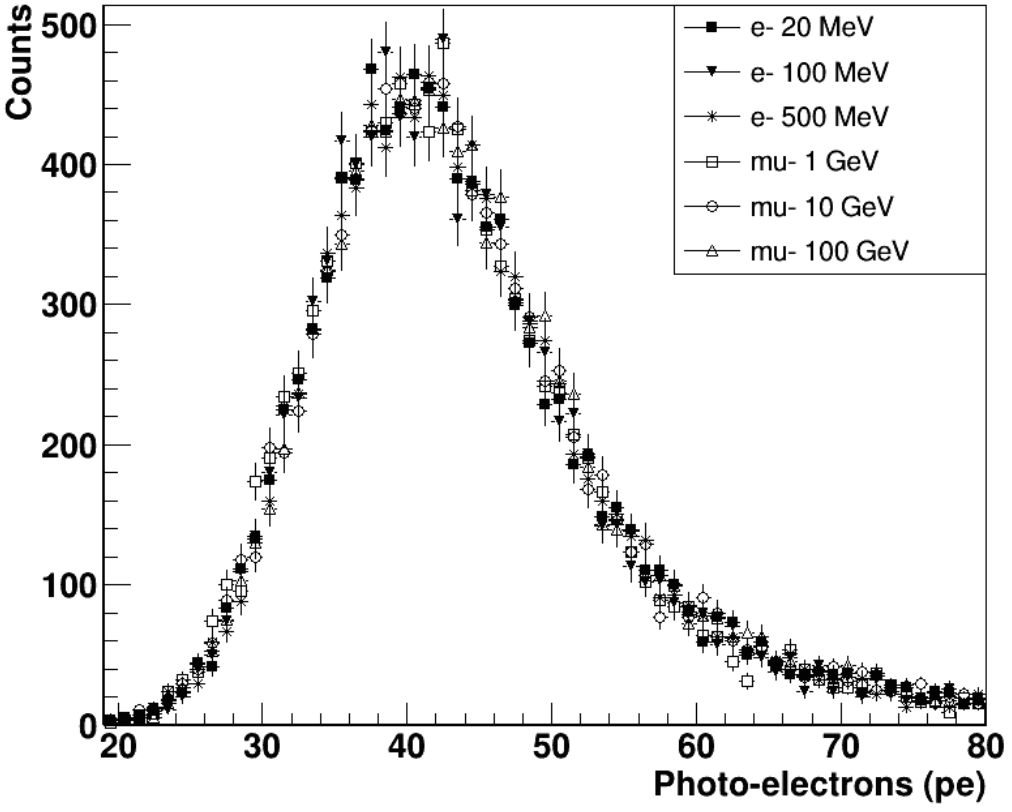}
    \caption{Scintillator bar response to charge particles of different energy. As expected, both electrons and muons generate the same distribution profile, since the energy deposited by all those particles in polystyrene is $\sim 2.08$~MeV. To obtain these distributions, we performed the interaction of $10^4$ particles (of each energy and type) with the bar. The average number of photo-electrons is around $40$ \textsl{pe} for all the particles, i.e., this detector is not able to distinguish muon events from background.}
    \label{fig:mips}
\end{figure}

Figure \ref{fig:mips} displays the simulation results of the scintillator bar response to charged particles of different energies. The distribution of the number of photo-electrons generated by $10^4$ electrons of $20$~MeV, $100$~MeV and $500$~MeV has the same profile as those corresponding to the $10^4$ muons of $1$~GeV, $10$~GeV and $100$~GeV, with a mean value around $40$ \textsl{pe}. This occurs because both these particles have a similar stopping power in polystyrene, i.e.  $\frac{dE}{d\varrho_\mathrm{pol}}\approx 2 \,\text{MeV cm}^2\text{/g}$\,\cite{MichaelEtal2008}, so they all deposit about $2.08$~MeV of energy when passing through $1$~cm of polystyrene,
\begin{equation}
\label{stopingeq}
E_\mathrm{d} = 2 \,\text{MeV cm}^2\text{/g} \times \varrho_\mathrm{pol} = 2 \,\text{MeV cm}^2\text{/g} \times l_\mathrm{pol} \times \rho_\mathrm{pol} \approx  \text{2.08} \, \text{MeV},
\end{equation}
where $l_\mathrm{pol}=1$ cm, is the distance traveled by the particles in the bar and $\rho_\mathrm{pol}=1.04$ g/cm$^3$ is the polystyrene density. Therefore this detector is not able to distinguish muons from electrons, and it is necessary to use the \textsl{WCD} to distinguish between muon events and background. 

\subsubsection{SiPM-Fiber coupling}
\begin{figure}
    \centering
    \includegraphics[scale=0.36]{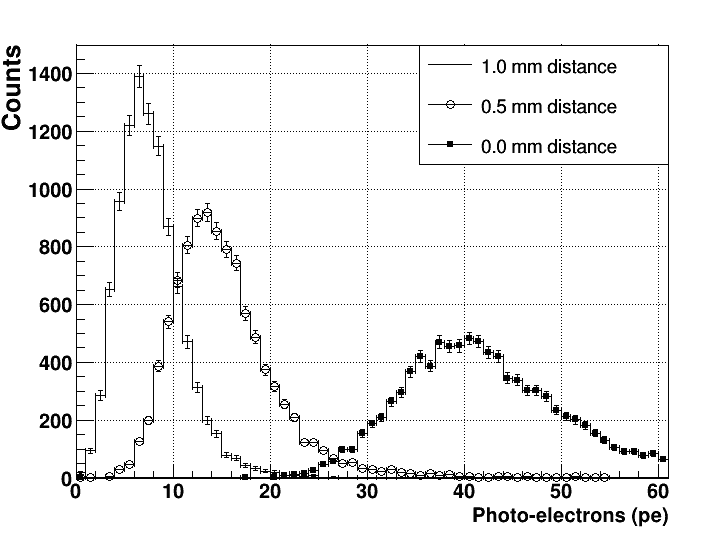}\,\,\,\includegraphics[scale=0.2]{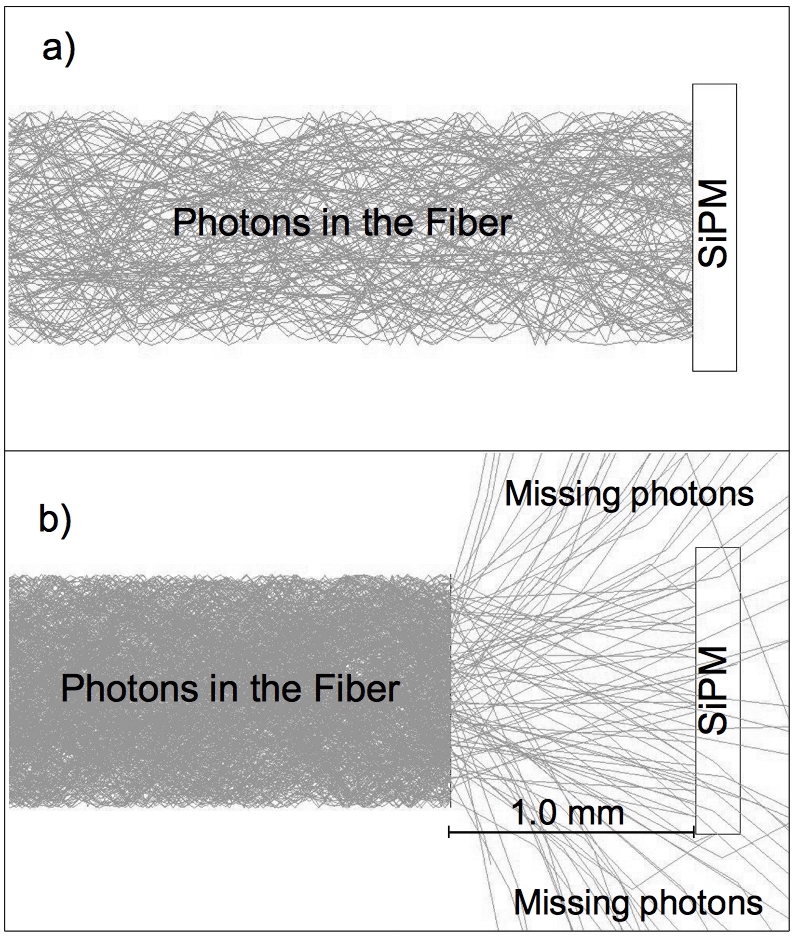}
    \caption{Distribution of the number of \textsl{pe} resulting from the evaluation of the \textsl{SiPM} and fiber coupling. a) The simulation of the ideal case, when the \textsl{SiPM} and the fiber are side by side. In this case, the average number is around $40$ \textsl{pe} and decreases with increasing distance between the \textsl{SiPM} and fiber. b) At $1.0$ mm the average number is around $7$ \textsl{pe}, i.e. $82$\% of the signal is lost.}
    \label{fig:coupling}
\end{figure}

The \textsl{Geant4} model allows estimating the quality of the coupling between \textsl{SiPM} and fiber. This coupling is ideal when they are side by side, without any space between them (see figure \ref{fig:coupling} plate \textit{a}). Thus, all the travelling photons at the edge of the fiber impact directly on the \textsl{SiPM} obtaining an average number of photo-electrons for an ideal coupling of $40$~\textsl{pe}. Two non-ideal coupling were simulated with distances of 0.5 mm and 1.0 mm between the \textsl{SiPM} and the fiber (shown in figure \ref{fig:coupling} plate \textit{b}). This space is filled with air to make the simulation as real as possible. The number of \textsl{pe} reduces to $7$ at $1$~mm of distance, representing a loss of 82\% of the signal compared to the ideal case, as shown in figure \ref{fig:coupling}. This result gives an idea of when a \textsl{Bar}-\textsl{Fiber}-\textsl{SiPM} system of the real detector is poorly coupled and should be replaced/checked.
\subsubsection{Attenuation of the photons in the \textsl{Bar}-\textsl{Fiber}-\textsl{SiPM} system}
\label{attenuation}
The light that propagates within the \textsl{WLS} fiber suffers an inevitable attenuation due to: photons escaping from the optical guide, and others absorbed by the material while being transported to the \textsl{SiPM}. 

\begin{figure}
    \centering
    \includegraphics[scale=0.3]{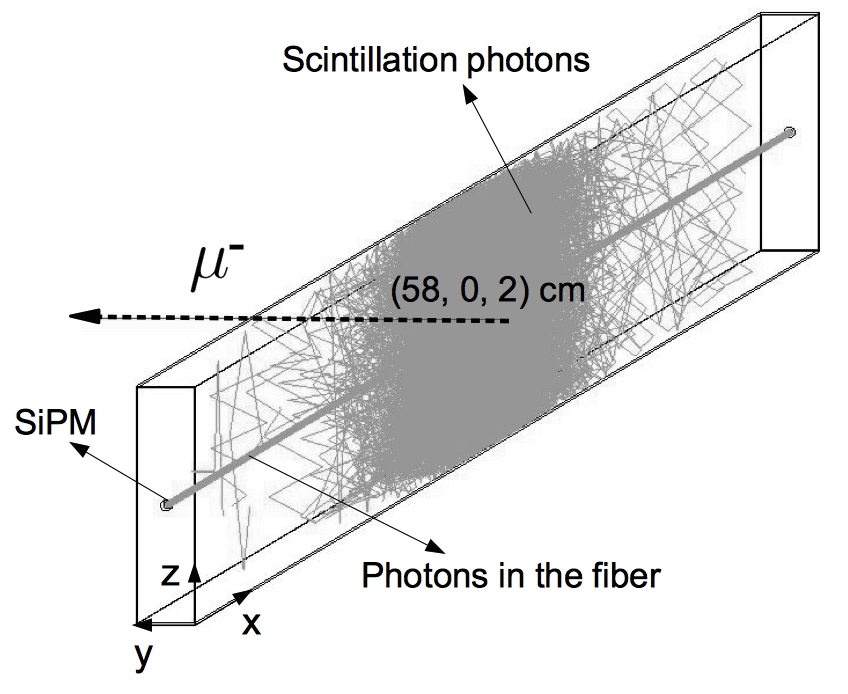}
    \caption{A muon (dash line) impinging the bar model at the point ($58$~cm, $0$~cm, $2$~cm) from the SiPM, producing scintillation photons that are absorbed and re-emitted while travelling within the fiber. These photons produce photo-electrons in the \textsl{SiPM} located at one end of the fiber. This simulation was performed with \textsl{MRS}s hitting at different $x$-positions to evaluate the attenuation of the \textsl{Bar}-\textsl{Fiber}-\textsl{SiPM} system.}
    \label{fig:event_bar}
\end{figure}

From the simulations of the scintillation detector, we can count the number of \textsl{pe} generated in the \textsl{SiPM} when a \textsl{MRS} crosses the bar at different distances, $x$, from the point of impact to the \textsl{SiPM} location (see Figure \ref{fig:event_bar}). These distances were chosen to take into account the width of the pixels of the hodoscope, i.e., $4$~cm wide. Therefore $x$ varies as, $x = (2+4p)$~cm,  $p=0,1,2,...,29$, and we obtain a distribution of the number of photo-electrons produced in SiPM as a function of $x$. We average its value  through a Gaussian adjustment. The attenuation values are normalized to the maximum value, i.e. $40$ \textsl{pe} (when the \textsl{MRS} impacts at $2$~cm from the SiPM). 

\begin{figure}
    \centering
        \includegraphics[scale=0.5]{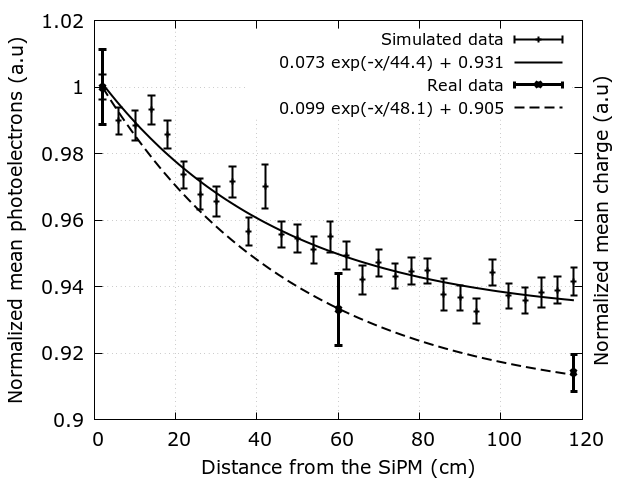}
   \caption{The normalized mean value of photoelectrons \textit{vs} the bar impacting position of the \textsl{MRS}. The maximum value is at $x=2$~cm, and as the particle impacts farther away from the \textsl{SiPM}, the intensity decreases by $\sim 7$\% in the simulated data and by $\sim 9$\% in the real data. Two exponential functions fit the simulated and the real data with attenuation factors, $\tau_s = 44.4$ and $\tau_r = 48.1$, that are almost the same. This behaviour can be associated with the attenuation of the photons travelling through the fiber.}
   \label{atenuacion_barra}
\end{figure}

Figure \ref{atenuacion_barra} shows the result of this simulation. If the particle impacts at the position closest to the \textsl{SiPM} (i.e. $x=2$~cm), the simulated data has the maximum mean value of photo-electrons that decreases with the distance from the \textsl{SiPM} as an exponential function $F(x)$,
\begin{equation}
F(x)= \text{0.073}\exp{\left( -\frac{x}{\tau_s}\right)} + 0.931.
\end{equation}
From this plot, it follows that the attenuation in the \textsl{Bar}-\textsl{Fiber}-\textsl{SiPM} system is around $7$\%, which agrees with the experimental data ($\sim9\%$) as shown in figure \ref{atenuacion_barra}. The fit function of the attenuation in the measured data is,
\begin{equation}
G(x)= \text{0.099}\exp{\left( -\frac{x}{\tau_r}\right)} + 0.905,
\end{equation}
and the comparison with $F(x)$ shows a resemblance between both attenuation factors, $\tau_s = 44.4$ and $\tau_r = 48.1$, which indeed are very close. In section \ref{sec:bar-data} we discuss the details of the experimental set-up to obtain the measured attenuation.

\begin{figure}
    \centering
    \includegraphics[scale=0.8]{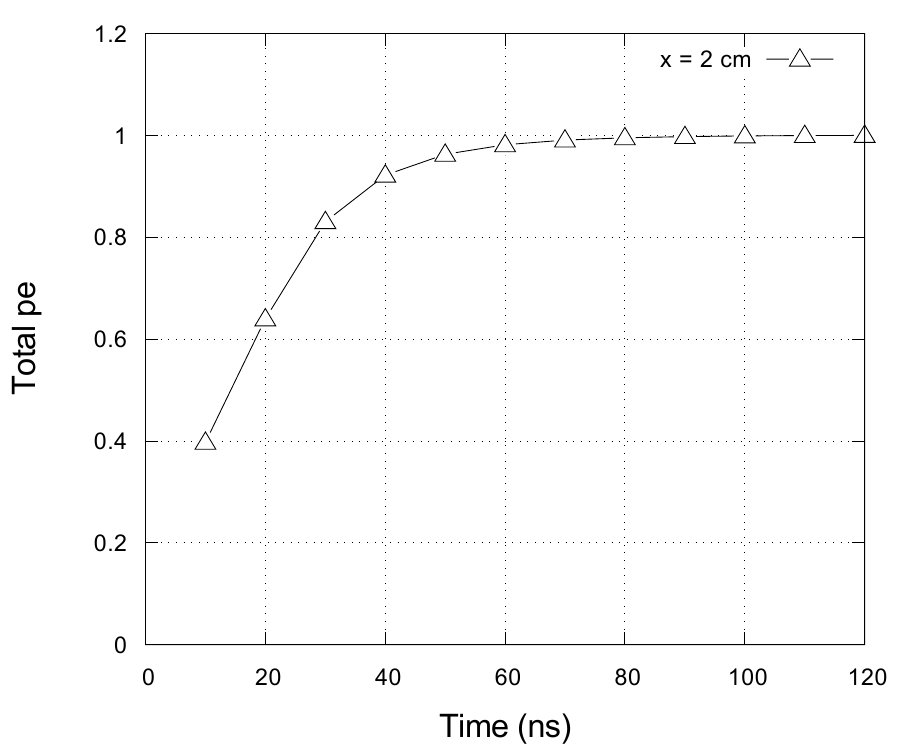}
    \includegraphics[scale=0.8]{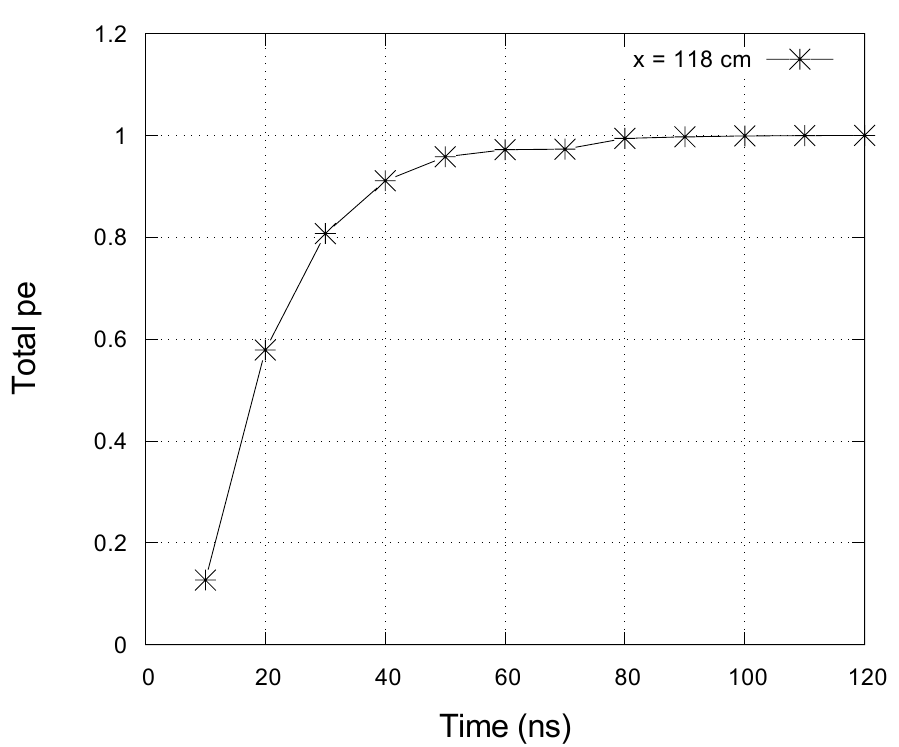}
    \caption{Cumulative number of photo-electrons produced when a \textsl{MRS} hits the bar at $2$~cm from the  \textsl{SiPM} (left) and at $118$~cm (right). It can be noticed that $40$\% of the Total \textsl{pe} occur in the first $10$ ns when $x=2$~cm. When $x=118$~cm only $12$\% of the \textsl{pe} comes in about the same time. In both cases, the total number of \textsl{pe} is reached in about $80$ ns, which emerges as the average time necessary to collect the total of \textsl{pe} produced by the passage of a muon, at any point of impact in the bar.}
    \label{cumulative}
\end{figure}

The \textsl{MRS} detection with scintillator bar generates a number of \textsl{pe} in the \textsl{SiPM} at a time $t$ and we can use figure \ref{cumulative} to estimate the time needed to collect the total number of photo-electrons produced. From the left plate of figure \ref{cumulative} we notice that $40$\% of the total photo-electrons occurs in the first $10$ ns when the \textsl{MRS}s hit the bar $2$~cm from the \textsl{SiPM}. From the right plate of the same figure, we observe that only $12$\% of the \textsl{pe} are produced in the same time, when the \textsl{MRS}s impact at $118$~cm from the \textsl{SiPM}. In both cases, the total number of \textsl{pe} is reached by $80$ ns, which is, the average time necessary to collect the total of \textsl{pe} produced by a muon impinging at any point of impact in the bar.

\subsection{Simulated attenuation in the hodoscope}
\label{sec:hodoscope-response-two}
\begin{figure}
    \centering
        \includegraphics[scale=0.5]{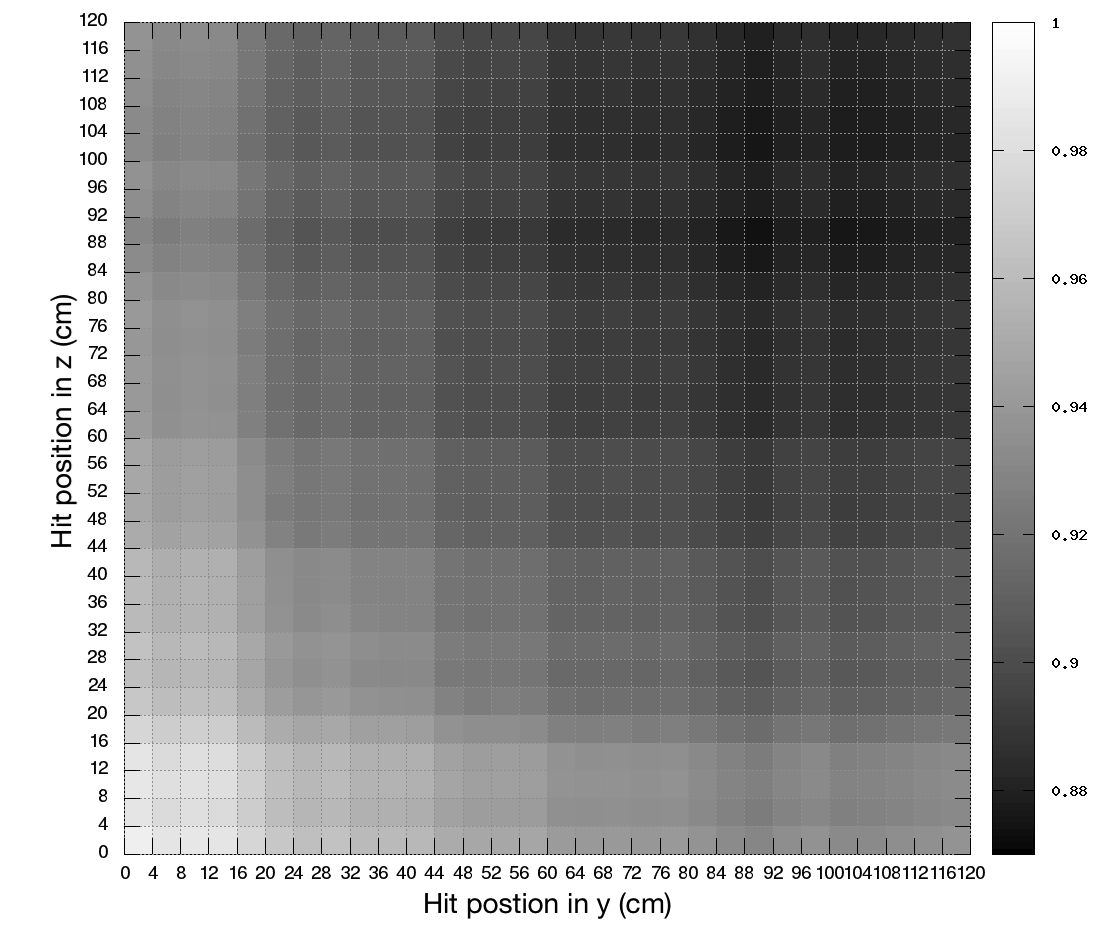}
   \caption[The response of the hodoscope panels]{The probability of photo-electron production in each pixel of a hodoscope panel by muons of $3$~GeV of energy. Each frame represents a detection pixel and as it can be seen, there is a gradual decrease in y and z-direction, as the pixel is further away from the SiPM. In pixel $P_{11}$, where the incoming muon is closest to the \textsl{SiPM}s, the maximum number of \textsl{pe} are produced; the number decreasing with distance up to 12\% at the furthest position. This result is valid for both front and rear panels.}\label{atenuacion_panel_bw}
\end{figure}
The probability of producing photo-electrons in the horizontal and vertical scintillators bar is a crucial concept in determining the response of each panel of the hodoscope. The bar simulation generates the response of a detection panel to \textsl{MRSs}, and due to attenuation, independent events in a particular pixel $P^{F}_{i,j}$ are given by 
\begin{equation}
\label{pe_panel}
P^{F}_{i,j}=P^{F}_{i} \times P^{F}_{j},
\end{equation}
where $P^{F}_{i}$ and $P^{F}_j$ represent the probabilities from the horizontal ($i$) and vertical ($j$) bars, respectively.

Figure \ref{atenuacion_panel_bw} displays the probability of photo-electrons produced by $3$~GeV muons striking each panel pixel, obtained from equation \ref{pe_panel}. These results are valid for the front and rear panels, and display a gradual decrease in both $y$ and $z$-direction, as the pixel is further away from the SiPM. In pixel, $P_{11}$ the maximum number of \textsl{pe} is produced and then decreases with distance up to $12$\% at the furthest position. This difference can be associated with the attenuation of photons in the \textsl{Bar}-\textsl{Fiber}-\textsl{SiPM} system.

\subsection{Experimental set-up to measure the \textsl{Bar}-\textsl{Fiber}-\textsl{SiPM} signal attenuation}
\label{sec:bar-data}
\begin{figure}
    \centering
        \includegraphics[scale=0.35]{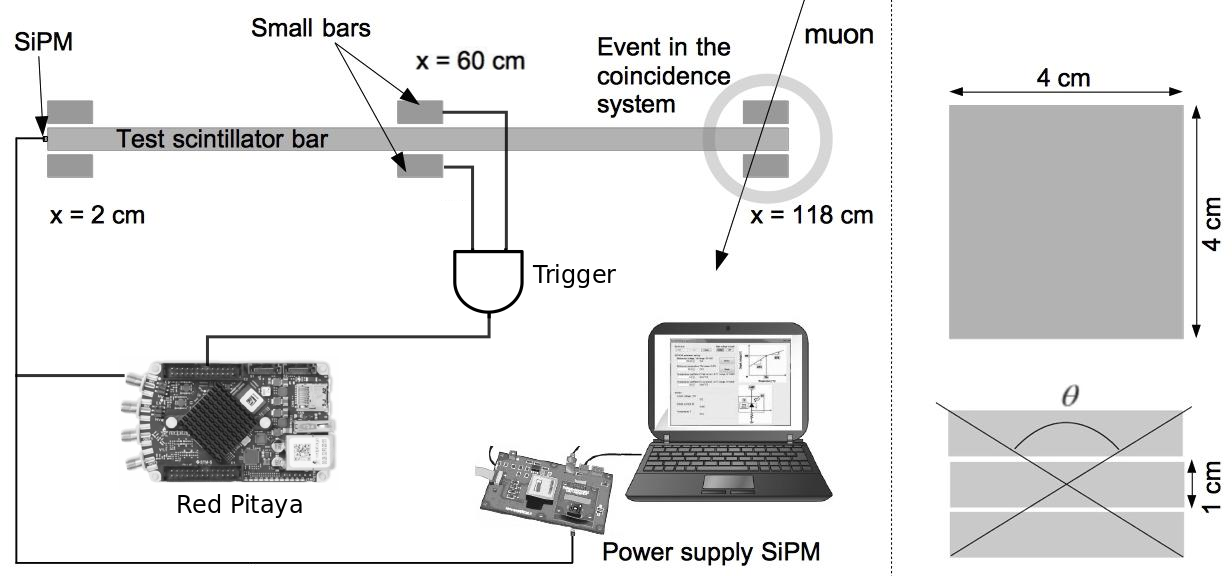}
   \caption{Diagram of the experimental setup to measure the \textsl{Bar}-\textsl{Fiber}-\textsl{SiPM} signal attenuation. Confirmed event emerges from a simultaneous signal from the coincidence system, i.e. the small upper bar, the test bar, and the lower bar. The dimensions of small bars, are shown, as well as the maximum angle $\theta = 106.26^{\circ}$ of the configuration. The data were collected at different particle impact distances from the \textsl{SiPM} ($2$~cm, $60$~cm and $118$~cm).} 
   \label{coincidencia_barras}
\end{figure}
Figure \ref{coincidencia_barras} displays the experimental setup of the coincidence system. An event is recorded if a simultaneous signal appears in the scintillator bar and in two small bars above and below. The maximum angle for the coincidence system is $\theta = 106.26^{\circ}$. To estimate the attenuation of the \textsl{Bar}-\textsl{Fiber}-\textsl{SiPM} system, we measure the signal produced by charged particles crossing at three points: $x=2$~cm, $x=60$~cm and $x=118$~cm. 

The acquisition system (Red Pitaya development board) records pulses with a frequency of $125$ MHz. The system was synchronized for coincidence pulses at each position in the test scintillator bar. The frequency of events per minute, measured with an oscilloscope Tektronix TDS 2002B, was $10\pm1$ per minute for $16$~cm$^2$. Thus, with this rate, it is possible to detect $0.625$~particles~min$^{-1}$~cm$^{-2}$. After calibrating for noise, we recorded $10^4$ pulses in the three positions. From each averaged pulse, the mean deposited charge was estimated by calculating the attenuation percentage, and normalizing values using figure \ref{atenuacion_barra}. 

The thermal stability of the \textsl{SiPM}s was analyzed before performing the measurements. The \textsl{SiPM} can vary their operation mode when the bias voltage, $V_{Bias}$, is less than the breakdown voltage, $V_B$, which changes with temperature. Therefore, a temperature-controlled box regulated by a TEC1-12706 thermoelectric Peltier cell was built and mounted on an aluminium frame, where we placed the \textsl{SiPM}. 

\begin{figure} 
    \centering
        \includegraphics[scale=0.6]{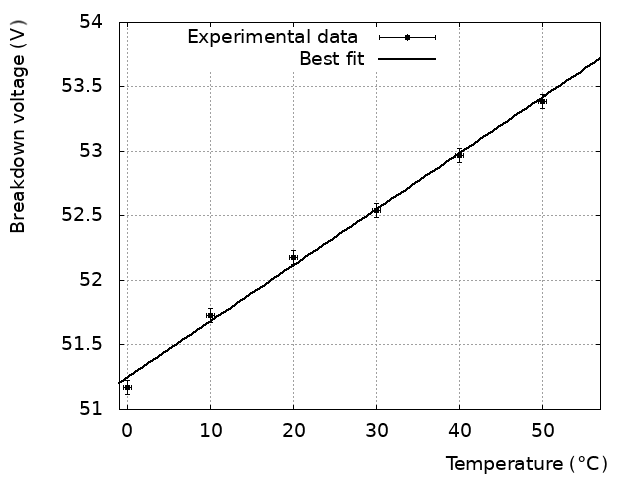}
   \caption{Dependence of the breakdown voltage with the temperature for the \textsl{SiPM} Hamamatsu S13360-1350CS used in the \textsl{MuTe} hodoscope. It can be observed that the relation is linear, that is for every $10^{\circ}$C of temperature the $V_B$ varies about $0.45$ V.}\label{temperature}
\end{figure}

Dark-Current\,\cite{Renker2006} measurements ranging from $0^{\circ}$C to $50^{\circ}$C illustrate the linear  dependence of $V_B$ with temperature (see figure \ref{temperature}). It is clear that with each $10^{\circ}$C of temperature, the $V_B$ changes around $0.45$~V.

\section{Modelling \textsl{MuTe} water Cherenkov detector response}
\label{sec:wcd-response}

The \textsl{WCD} indirectly detects charged particles, by the Cherenkov light emission from those secondaries travelling through the water. The photo-multiplier tube detects photo-electrons according to its quantum efficiency, which depends on the wavelength of the impacting photon. Next, we will discuss some results from the detector simulation, as well as the first data recorded by the \textsl{WCD} in Bucaramanga, Colombia.
\subsection{The  \textsl{WCD} model}
\label{sec:wcd-model}
The \textsl{WCD} \textsl{Geant4} model is a stainless steel cube of length $l_c=121$~cm, filled with purified water. The water has a refractive index $n$, which varies between $1.3435$ and $1.3608$, and a photon absorption length ranging from $0.69$~m to $2.90$~m depending on the energy. The Tyvek is a \textit{LUT} model between the water and the container, with \textit{groundtyvek} finish, a \textit{dielectric-LUT} interface and a reflection index $n_{\mathrm{Tyvek}} = 1$. 

\begin{figure}
    \centering
    \includegraphics[scale=0.3]{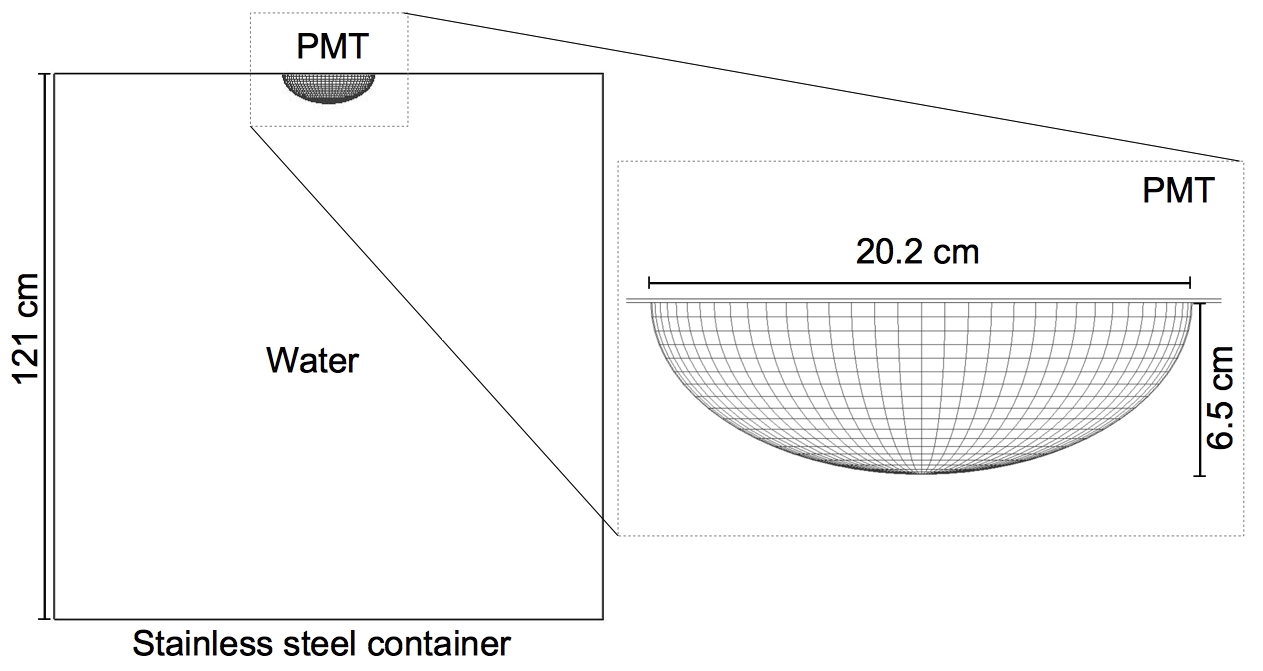}
    \caption{Geant4 model of the WCD. The container is made of stainless steel, with the inside walls covered with Tyvek, and modelled as an optical surface between the container and water, with a reflection index $n_{\mathrm{Tyvek}} = 1$. The PMT model is a half-ellipsoid full of air.}
    \label{fig:wcd_dimentions}
\end{figure}

We simulate the photo-cathode of the \textsl{PMT}  as a half-ellipsoid located on top of the water cube, as shown in figure \ref{fig:wcd_dimentions}. The Quantum Efficiency (\textsl{QE}) function of this device was introduced in the code, taking into account the reference data of the Hamamatsu R5912\,\cite{Hamamatsu2018}, and represents an improvement over previous simulations of \textsl{WCD}s \cite{CalderonAsoreyNunez2015}. The \textsl{QE} determines whether photons reaching the outer surface of the photo-cathode will produce photo-electrons. The \textsl{WCD} response is given in terms of the number of \textsl{pe} generated by each particle interacting with it.

 \subsection{Estimation of the Vertical Equivalent Muon unit}
The Vertical Equivalent Muon (\textsl{VEM}) --defined as the average charge collected in the \textsl{PMT} when a high-energy muon crosses the entire detector vertically--  is generally adopted as the unit in the calibration of the energy deposited by incident particles, and is independent of the detection position\,\cite{EtchegoyenEtal2005}.

The \textsl{Geant4} simulation allows the injection of muons with a chosen energy and direction. In the modelling, we inject $10^{5}$ vertical muons with $3$~GeV over the \textsl{WCD}. These particles produce many Cherenkov photons, $N$, which strike the surface of the photo-cathode, producing numerous photo-electrons, $N_{\mathrm{pe}}$, depending on their wavelength. We can estimate the efficiency of the \textsl{WCD} in the following way: one \textsl{VEM} of $3$~GeV generates around $46857$ Cherenkov photons in $120$~cm of water. Since only $1617$ of those photons reach the external surface of the photo-cathode, and on average about $203.2$ \textsl{pe} are produced, the system has a muon detection efficiency of $0.4$\%, that is,
\begin{equation}
\eta_{\mathrm{WCD}}=\frac{N_{\mathrm{FE}}}{N} 100\%=\frac{203.2}{46857}100\%=0\textnormal{.}4\%
\end{equation}

\begin{figure}
    \centering
    \includegraphics[scale=0.45]{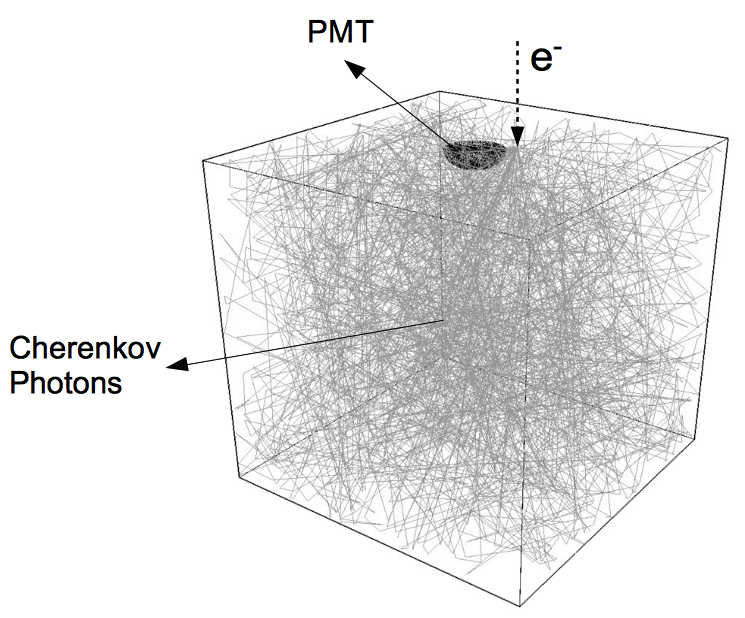}
    \includegraphics[scale=0.3]{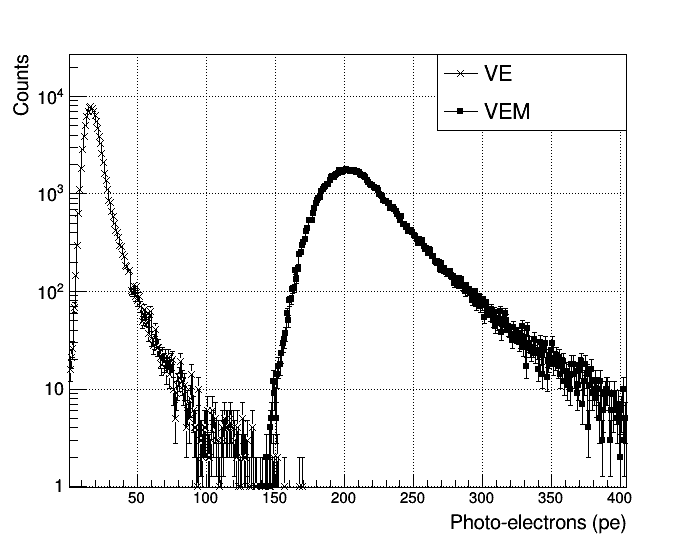}
    \caption{ On the left plate we illustrate Cherenkov photons produced inside the \textsl{WCD}. In the right plate we present the distribution of the number of photo-electrons produced due to the detection of vertical muons (squares) and vertical electrons (crosses) with the \textsl{WCD}. The mean of the \textsl{VEM} curve is $\sim 203.2$ \textsl{pe} compared to $16.7$ \textsl{pe} for \textsl{VE}, that is the $8$\% of the \textsl{VEM}.}
    \label{vem_ve}
\end{figure}

To compare the \textsl{WCD} response to muons and electrons at Bucaramanga, we simulated the impact of $10^{5}$ vertical electrons (\textsl{VE}) of $20$~MeV with the same initial directions and impact points (see figure \ref{vem_ve}). The histograms illustrating the distribution of the number of photo-electrons of \textsl{VE} and \textsl{VEM} are shown in figure \ref{vem_ve}, where we can see that the mean value of the \textsl{VE} is smaller than that of the \textsl{VEM}, $\sim 16.7$\textsl{pe}, i.e the \textsl{VE} is around $\sim 8$\% of the \textsl{VEM}.
 
\begin{figure}
    \centering
    \includegraphics[scale=0.3]{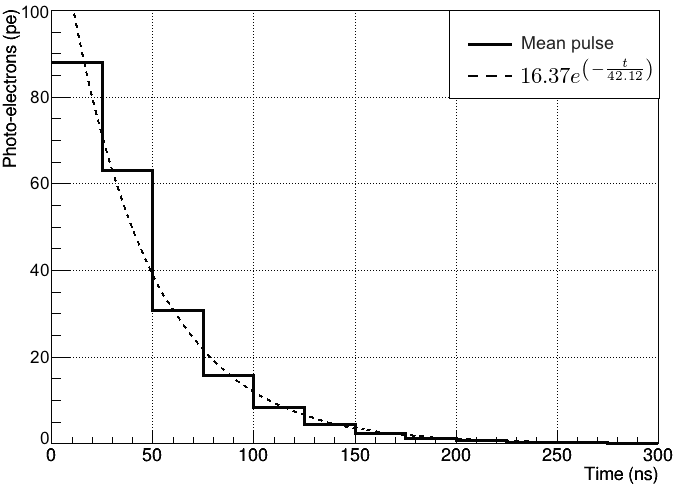}
    \includegraphics[scale=0.3]{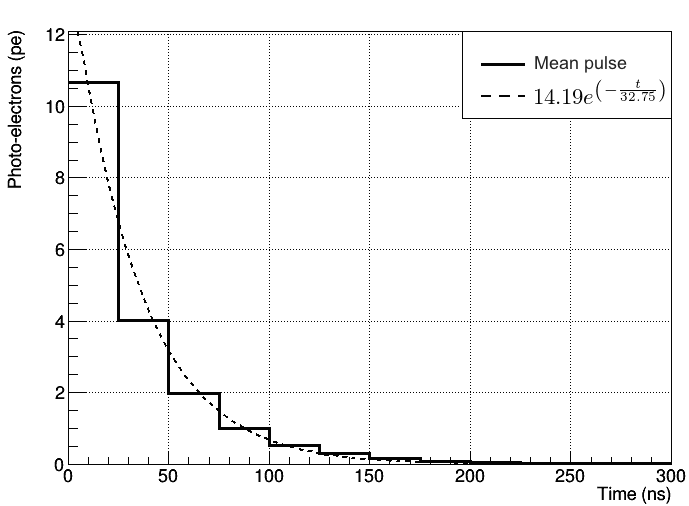}
    \caption{Mean pulse corresponding to the \textsl{VEM} (left) and the \textsl{VE} (right) response. The dashed lines represent the best exponential fit of the pulses where the attenuation time $\tau$ is around $42.12$ ns for the \textsl{VEM} and $32.75$ ns for the \textsl{VE}.}
    \label{pulse_vem_ve}
\end{figure}

Next, in figure \ref{pulse_vem_ve} the number of \textsl{pe} versus time is shown for both pulses: \textsl{VEM} (left) and \textsl{VE} (right).  We obtain attenuation in time and length from the best fits to the distribution histograms. Finally,  table 1 summarizes the comparison between both muons and electrons.

\begin{table}
\label{comparacion}
\centering
\caption{Summary of the physical magnitudes obtained for \textsl{VEM} and \textsl{VE}: Length traveled in water ($l$), Number of Cherenkov Photons produced ($N$), Number of Photons that reach the \textsl{PMT} ($N_{\mathrm{PMT}}$), Number of Photoelectrons ($N_{\mathrm{pe}}$), Time of attenuation of the pulse ($\tau $) and Length of attenuation ($l_a$).}
\begin{tabular}{l|c|c|}
\cline{2-3}
                                                     & \textbf{$\mu^-$ ($3$~GeV) }& \textbf{$e^-$ ($20$~MeV)} \\ \hline
\multicolumn{1}{|l|}{\textbf{$l$}}   &          (120 $\pm$ 1) cm      &    (10 $\pm$ 1) cm        \\ \hline
\multicolumn{1}{|l|}{\textbf{$N$}}    &         46857 $\pm$ 13 $ $   &      3538 $\pm$ 1 $ $      \\ \hline
\multicolumn{1}{|l|}{$N_{\mathrm{PMT}}$}       &          1617 $\pm$ 1     &     132.1 $\pm$ 0.1      \\ \hline
\multicolumn{1}{|l|}{$N_{\mathrm{pe}}$}       &          203.2 $\pm$ 0.2      &     16.729 $\pm$ 0.003       \\ \hline
\multicolumn{1}{|l|}{$\tau$} &          (42.12 $\pm$ 0.01) ns      &      (32.75 $\pm$ 0.03) ns      \\ \hline
\multicolumn{1}{|l|}{$l_a$} &          (7.332 $\pm$ 0.001) m      &      (9.430 $\pm$ 0.002) m      \\ \hline
\end{tabular}
\end{table}
\subsection{\textsl{WCD} response to the cosmic ray background radiation}
\begin{figure}
    \centering
    \includegraphics[scale=0.5]{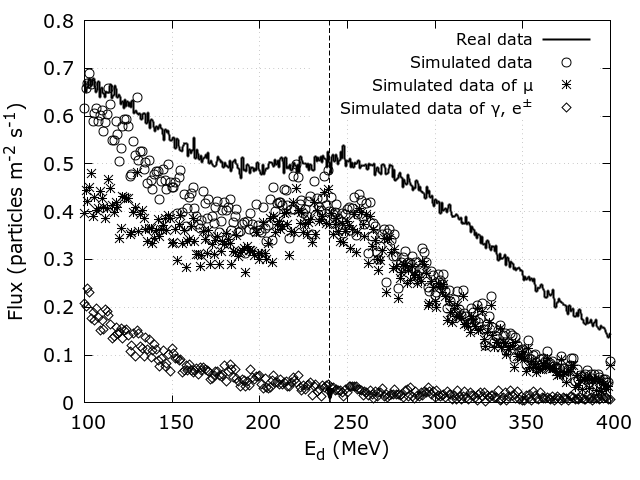}
    \caption{Energy deposited (simulated and detected) by particles in the \textsl{WCD}, at Bucaramanga level.  From the simulated data, it is possible to identify the preeminence of the muonic component to the second hump of the distribution (from $200$~MeV to $300$~MeV), and it coincides with the \textsl{VEM} simulated value (dashed line). Therefore we can discriminate the events corresponding to the muon component of the impacting particles.}
\label{energy_deposited_wcd}
\end{figure}
The \textsl{LAGO}-\textsl{ARTI} framework allows us to estimate the response of the \textsl{WCD} to the cosmic ray background radiation flux ($\Xi$) at any site\,\cite{SarmientoEtal2019}. This toolkit employs a \textsl{Geant4} code to estimate the number of Cherenkov photons detected by the \textsl{PMT} with its quantum efficiency. The \textsl{ARTI} framework uses the energy and the momentum of the particles from $\Xi$ as input and obtains the flux by using the \textsl{CORSIKA} code\,\cite{HeckEtal1998}, with a geomagnetic correction from the \textsl{MAGCOS} code\,\cite{Desorgher2003}. For  examples of this precise simulation chain see references \,\cite{AsoreyEtal2015B} and  \cite{AsoreyNunezSuarez2018}. 
 
Figure \ref{energy_deposited_wcd} compares the simulated and measured data of the energy deposited in the \textsl{WCD} by all particles at Bucaramanga\,\cite{PenarodriguezEtal2019}. From the simulated data, it is possible to identify the second hump of the distribution, dominated by the muonic component, i.e., from $200$~MeV to $300$~MeV we can discriminate the events corresponding to the muon component of the impacting particles.

\section{Conclusions}
\label{sec:conclusions}
The \textsl{MuTe} detector is an innovative setup which differentiates it from some previous detectors, combining two independent detectors:
\begin{itemize}
    \item \textbf{Scintillator hodoscope:} Inspired by the experiences of other volcano muography experiments\,\cite{UchidaTanakaTanaka2009,GibertEtal2010}, we have designed an hodoscope with two X-Y  arrays of $30 \times 30$ plastic scintillating strips leading to $900$ pixels of $4$~cm$ \, \times \, 4$~cm = $16$~cm$^2$, summing up $14.400$~cm$^2$ of detection surface which can be separated up to a distance of $250$~cm.
    \item \textbf{Water Cherenkov Detector:} The \textsl{WCD} is a purified water cube of $120$~ cm side, located behind the rear scintillator panel, which acts as an absorbing element and as a third active coincidence detector. 
\end{itemize}

From the \textsl{Geant4} modelling of the scintillator detector, we obtain that the number of \textsl{pe} decreases about $7$\%, with regard  to those produced at the near end of the \textsl{SiPM}. This reduction occurs due to the attenuation of the photons that travel in the fiber, since, the more distance they travel within it, the more energy they lose and thus fewer photons reach the \textsl{SiPM}s. This result is in agreement with the $9$\% attenuation obtained with the experimental setup described in section \ref{sec:bar-data}. This attenuation seems to be insignificant in the bar, but it is more noticeable in the hodoscope panels, since, the difference between the closest corner to the \textsl{SiPM}s and the furthest, is around $12$\%.

The results obtained lead to a definition of a \textsl{MuTe} muon detection trigger in terms of the energy deposited by each component. That is, muons must deposit about $2.08$~MeV in two scintillator bars on the front panel, then about the same energy in two bars on the rear panel of the hodoscope, and finally be discriminated from the background in the \textsl{WCD}. Lastly the muon must lose about $240$~MeV to be counted as an event.

 The capacity for determining an accurate muon signal emerging from a volcano is a critical factor for muography studies. There could be fake signals produced in the hodoscope by electrons/positrons because the signal produced in the hodoscope is equal for muon and electrons  as shown in figure \ref{fig:mips}. Due to the dimensions and location, the \textsl{WCD} filters most of the upward background (low energies electrons, protons, and muons coming from the backward free sky), which could cause overestimation in the hodoscope counts\,\cite{NishiyamaEtal2016}. It is also capable of isolating the muonic component of the incident particle flux, as shown in figure \ref{vem_ve}.  Additionally, from the charge histogram, obtained by time integration of the individual pulses measured in the \textsl{WCD}, it is possible to separate the two components of the incident flux: electromagnetic part (photon, electron \& positron) and the $\mu-$component\,\cite{AsoreyEtal2015B}.

The computational model developed and presented in this work is a first approximation in understanding the response of \textsl{MuTe} to the cosmic ray background, in view that this instrument works as an observatory. In the coming months, we will be installing the \textsl{MuTe} detector in Cerro Mach\'in volcano\,\cite{AsoreyEtal2017B} and start on-field data acquisition, providing new information about the density distribution inside the volcanic edifice.

\acknowledgments
We gratefully acknowledge the observations, sugestions and criticisms for the anonymous referees, improving the presentation and clarity of the present work. The authors recognise the financial support of  Departamento Administrativo de Ciencia, Tecnolog\'{i}a e Innovaci\'on de Colombia (ColCiencias) under contract FP44842-082-2015 and to the Programa de Cooperaci\'on Nivel II (PCB-II) MINCYT-CONICET-COLCIENCIAS 2015, under project CO/15/02.  We are also very thankful to LAGO and the Pierre Auger Collaboration for their continuous support.  The simulations in this work were partially possible thanks to The Red Iberoamericana de Computaci\'on de Altas Prestaciones (RICAP, 517RT0529), co-funded by the Programa Iberoamericano de Ciencia y Tecnolog\'{i}a para el Desarrollo (CYTED) under its Thematic Networks Call. We also thank the computational support from the Universidad Industrial de Santander (SC3UIS) High Performance and Scientific Computing Centre. We would also like to thank Vicerrector\'{i}a Investigaci\'on y Extensi\'on Universidad Industrial de Santander for its permanent sponsorship. We recognise many very fruitful discussions with D. Sierra-Porta, thanks a lot.


\begin{thebibliography}{10}

\bibitem{MarteauEtal2012}
J.~Marteau, et al.
\newblock Muons tomography applied to geosciences and volcanology.
\newblock {\em Nucl. Instrum. Meth. A}, 695:23 -- 28, 2012.
\newblock New Developments in Photodetection NDIP11.

\bibitem{Kaiser2019}
R.~Kaiser.
\newblock {Muography: overview and future directions}.
\newblock {\em Philos. T. R. Soc. A.}, 377(2137):20180049, January 2019.

\bibitem{BlanpiedEtal2015}
G.~Blanpied, et al.
\newblock Material discrimination using scattering and stopping of cosmic ray
  muons and electrons: Differentiating heavier from lighter metals as well as
  low-atomic weight materials.
\newblock {\em Nucl. Instrum. Meth. A}, 784:352--358, jun 2015.

\bibitem{MorishimaEtal2017}
K.~Morishima, et al.
\newblock Discovery of a big void in khufu's pyramid by observation of
  cosmic-ray muons.
\newblock {\em Nature}, 552(7685):386, 2017.

\bibitem{GomezEtal2016}
H.~G\'omez, et al.
\newblock Studies on muon tomography for archaeological internal structures
  scanning.
\newblock In {\em J. Phys. Conf. Ser.}, page 052016. IOP Publishing, 2016.

\bibitem{FujiiEtal2013}
H.~Fujii, et al.
\newblock Performance of a remotely located muon radiography system to identify
  the inner structure of a nuclear plant.
\newblock {\em Prog. Theor. Exp. Phys.}, 2013(7), jul 2013.

\bibitem{SaracinoEtal2017}
G.~Saracino, et al.
\newblock Imaging of underground cavities with cosmic-ray muons from
  observations at mt. echia (naples).
\newblock {\em Sci. Rep.}, 7(1), apr 2017.

\bibitem{ThompsonEtal2019}
L.~F. {Thompson}, et al.
\newblock {The application of muon tomography to the imaging of railway
  tunnels}.
\newblock {\em arXiv e-prints}, page arXiv:1906.05814, Jun 2019.

\bibitem{TanakaOlah2019}
H.~K.~M. Tanaka and L.~Ol{\'a}h.
\newblock {Overview of muographers}.
\newblock {\em Philos. T. R. Soc. A.}, 377(2137):20180143, January 2019.

\bibitem{Cortes2016}
G.P. Cort{\'e}s.
\newblock Informe de actividad volc\'anica segmento norte de colombia diciembre
  de 2016.
\newblock Technical report, Reporte interno, Manizales, Colombia. INGEOMINAS,
  2016.

\bibitem{Agudelo2016}
A.~Agudelo.
\newblock Informe t\'ecnico de actividad de los volcanes nevado del huila,
  purac\'e y sotar\'a, durante el periodo de diciembre de 2016.
\newblock Technical report, Reporte Interno, Popayan, Colombia, Servicio
  Geol\'ogico Colombiano, 12 2016.

\bibitem{Munoz2017}
E.~Mu{\~n}oz.
\newblock Informe mensual de actividad de los volcanes galeras,cumbal,chiles y
  cerro negro, las \'animas, dona juana y azufral.
\newblock Technical report, Reporte interno, Pasto, Colombia. INGEOMINAS, 2017.

\bibitem{AsoreyEtal2017B}
A.~Vesga-Ramirez, et al.
\newblock {Muon Tomography sites for Colombia volcanoes}.
\newblock {\em ArXiv e-prints, arXiv 1705.09884}, May 2017.

\bibitem{SierraPortaEtal2018}
H.~Asorey, et al.
\newblock Astroparticle projects at the eastern colombia region: facilities and
  instrumentation.
\newblock {\em Scientia et technica}, 23(3):391--396, 2018.

\bibitem{PenaRodriguezEtal2018}
H.~Asorey, et al.
\newblock minimute: A muon telescope prototype for studying volcanic structures
  with cosmic ray flux.
\newblock {\em Scientia et technica}, 23(3):386--390, 2018.

\bibitem{GuerreroEtal2019}
I.D. Guerrero, et al.
\newblock Design and construction of a muon detector prototype for study the
  galeras volcano internal structure.
\newblock In {\em J. Phys. Conf. Ser.}, page 012020. IOP Publishing, 2019.

\bibitem{ParraAvila2019}
J.S. Useche-Parra and C.A. Avila-Bernal.
\newblock Estimation of cosmic-muon flux attenuation by monserrate hill in
  bogota.
\newblock {\em J. Instrum.}, 14(02):P02015, 2019.

\bibitem{PenarodriguezEtal2019}
J.~Pe{\~n}a-Rodr{\'\i}guez, et al.
\newblock Calibration and first measurements of MuTe: a hybrid muon telescope
  for geological structures.
\newblock In {\em 36th International Cosmic Ray Conference (ICRC2019)},
  volume~36, 2019.

\bibitem{AnastasioEtal2013}
A.~Anastasio, et al.
\newblock The mu-ray detector for muon radiography of volcanoes.
\newblock {\em Nucl. Instrum. Meth. A}, 732:423 -- 426, 2013.
\newblock Vienna Conference on Instrumentation 2013.

\bibitem{CarloganuEtal2013}
C.~Carloganu, et al.
\newblock Towards a muon radiography of the puy de d{\^o}me.
\newblock {\em Geosci. Instrum. Meth.}, 2:55--60, 2013.

\bibitem{LesparreEtal2010}
N.~Lesparre, et al.
\newblock Geophysical muon imaging: feasibility and limits.
\newblock {\em Geophys. J. Int.}, 183(3):1348--1361, 2010.

\bibitem{Nagamine2016}
K.~Nagamine.
\newblock Radiography with cosmic-ray and compact accelerator muons; exploring
  inner-structure of large-scale objects and landforms.
\newblock {\em P. Jpn. Acad. B-Phys.}, 92(8):265--289, 2016.

\bibitem{SehgalEtal2016}
R.~Sehgal, et al.
\newblock Simulations and track reconstruction for muon tomography using
  resistive plate chambers.
\newblock In {\em DAE Symp. Nucl. Phys.}, volume~61, pages 1034--1035, 2016.

\bibitem{Fehr2012}
F.~Fehr and {Tomuvol Collaboration}.
\newblock Density imaging of volcanos with atmospheric muons.
\newblock In {\em J. Phys. Conf. Ser.}, page 052019. IOP Publishing, 2012.

\bibitem{BouteilleEtal2016}
S.~Bouteille, et al.
\newblock A micromegas-based telescope for muon tomography: The watto
  experiment.
\newblock {\em Nucl. Instrum. Meth. A}, 834:223--228, 2016.

\bibitem{OlahEtal2018}
L.~Ol{\'a}h, et al.
\newblock High-definition and low-noise muography of the sakurajima volcano
  with gaseous tracking detectors.
\newblock {\em Sci. Rep.}, 8(1):3207, 2018.

\bibitem{LesparreEtal2012}
N.~Lesparre, et al.
\newblock {Design and operation of a field telescope for cosmic ray geophysical
  tomography}.
\newblock {\em Geosci. Instrum. Meth.}, 1:33--42, 2012.

\bibitem{TanakaEtal2009}
H.K.M. Tanaka, et al.
\newblock Cosmic-ray muon imaging of magma in a conduit: Degassing process of
  satsuma iwojima volcano, japan.
\newblock {\em Geophys. Res. Lett.}, 36(1), 2009.

\bibitem{NagamineEtal1995}
K.~Nagamine, et al.
\newblock Method of probing inner-structure of geophysical substance with the
  horizontal cosmic-ray muons and possible application to volcanic eruption
  prediction.
\newblock {\em Nucl. Instrum. Meth. A}, 356(2):585 -- 595, 1995.

\bibitem{AguiarEtal2015}
P.~Aguiar, et al.
\newblock Geant4-gate simulation of a large plastic scintillator for muon
  radiography.
\newblock {\em IEEE Trans. Nucl. Sci.}, 62(3):1233--1238, 2015.

\bibitem{TangEtal2016}
S.W Tang, et al.
\newblock A large area plastic scintillation detector with 4-corner-readout.
\newblock {\em Chinese Physics C}, 40(5):056001, 2016.

\bibitem{AgostinelliEtal2003}
S.~Agostinelli, et al.
\newblock Geant4: a simulation toolkit.
\newblock {\em Nucl. Instrum. Meth. A}, 506(3):250--303, 2003.

\bibitem{NishiyamaMiyamotoNaganawa2014}
R.~Nishiyama, S.~Miyamoto, and N.~Naganawa.
\newblock Experimental study of source of background noise in muon radiography
  using emulsion film detectors.
\newblock {\em Geosci. Instrum. Meth.}, 3(1):29--39, April 2014.

\bibitem{KusagayaTanaka2015}
T.~Kusagaya and H.K.M. Tanaka.
\newblock Muographic imaging with a multi-layered telescope and its application
  to the study of the subsurface structure of a volcano.
\newblock {\em P. Jpn. Acad. B-Phys.}, 91(9):501--510, 2015.

\bibitem{NishiyamaEtal2016}
R.~Nishiyama, et al.
\newblock Monte carlo simulation for background study of geophysical inspection
  with cosmic-ray muons.
\newblock {\em Geophys. J. Int.}, 206(2):1039--1050, 2016.

\bibitem{GomezEtal2017}
H.~G{\'o}mez, et al.
\newblock Forward scattering effects on muon imaging.
\newblock {\em J. Instrum.}, 12(12):P12018, 2017.

\bibitem{SaintGobain2017}
Saint-Gobain Ceramics \& Plastics.
\newblock {\em Plastic Scintillating Fibers}, 2017.
\newblock Rev. 1.

\bibitem{Hamamatsu2018}
Hamamatsu.
\newblock {\em MPPCs for precision measurement}, 5 2018.
\newblock Rev. 1.

\bibitem{Filevich1999}
A.~Filevich, et al.
\newblock Spectral-directional reflectivity of tyvek immersed in water.
\newblock {\em Nucl. Instrum. Meth. A}, 423(1):108--118, 1999.

\bibitem{MichaelEtal2008}
D.G.~Michael, et al.
\newblock The magnetized steel and scintillator calorimeters of the minos
  experiment.
\newblock {\em Nucl. Instrum. Meth. A}, 596(2):190--228, 2008.

\bibitem{Pitaya2016}
Red Pitaya.
\newblock Red pitaya documentation.
\newblock {\em Release 0.97. url: http://redpitaya. readthedocs.
  io/en/latest/(visited on 02/11/2017)}, 2016.

\bibitem{Renker2006}
D. Renker.
\newblock Geiger-mode avalanche photodiodes, history, properties and problems.
\newblock {\em Nucl. Instrum. Meth. A}, 567(1):48--56, 2006.

\bibitem{CalderonAsoreyNunez2015}
R.~Calder{\'o}n, H.~Asorey, and L.A. N{\'u}{\~n}ez.
\newblock Geant4 based simulation of the water cherenkov detectors of the lago
  project.
\newblock {\em Nucl. Part. Phys. Proc.}, 267:424--426, 2015.

\bibitem{EtchegoyenEtal2005}
A.~Etchegoyen, et al.
  D.G. Melo, A.C. Rovero, A.D. Supanitsky, and A.~Tamashiro.
\newblock Muon-track studies in a water cherenkov detector.
\newblock {\em Nucl. Instrum. Meth. A}, 545(3):602 -- 612, 2005.

\bibitem{SarmientoEtal2019}
R.~Calder{\'o}n-Ardila, et al.
\newblock Modeling the LAGO's detectors response to secondary particles at
  ground level from the Antarctic to Mexico.
\newblock In {\em 36th International Cosmic Ray Conference (ICRC2019)},
  volume~36, 2019.

\bibitem{HeckEtal1998}
D.~Heck, et al.
\newblock Corsika : A monte carlo code to simulate extensive air showers.
\newblock Technical Report FZKA 6019, Forschungszentrum Karlsruhe GmbH, 1998.

\bibitem{Desorgher2003}
L.~Desorgher.
\newblock {MAGNETOSCOSMICS, Geant4 application for simulating the propagation
  of cosmic rays through the Earth magnetosphere}.
\newblock Technical report, Physikalisches Institut, University of Bern, Bern,
  Germany, 2003.

\bibitem{AsoreyEtal2015B}
H.~Asorey, et al..
\newblock The {LAGO} space weather program: Directional geomagnetic effects,
  background fluence calculations and multi-spectral data analysis.
\newblock In {\em The 34th International Cosmic Ray Conference}, volume
  PoS(ICRC2015), page 142, 2015.

\bibitem{AsoreyNunezSuarez2018}
H.~Asorey, L.~A. N{\'u}{\~n}ez, and M.~Su{\'a}rez-Dur{\'a}n.
\newblock Preliminary results from the latin american giant observatory space
  weather simulation chain.
\newblock {\em Space Weather}, 16(5):461--475, 2018.

\bibitem{UchidaTanakaTanaka2009}
T.~Uchida, H.~K.M. Tanaka, and M.~Tanaka.
\newblock Space saving and power efficient readout system for cosmic-ray muon
  radiography.
\newblock {\em IEEE Trans. Nucl. Sci.}, 56(2):448--452, 2009.

\bibitem{GibertEtal2010}
D.~Gibert, et al.
\newblock Muon tomography: Plans for observations in the lesser antilles.
\newblock {\em Earth Planets Space}, 62(2):153--165, 2010.

\end{thebibliography}

\end{document}